\documentclass[12pt,preprint]{aastex}
\usepackage{amsfonts}
\usepackage{amsmath}
\usepackage{amssymb}
\usepackage{amstext}

\begin{document}

\title{Gravitational Lensing of Supernovae Type Ia by Pseudo Elliptic NFW Haloes}

\author{Hamed Bagherpour %
\footnote{hamed@nhn.ou.edu } }

\affil{Homer L. Dodge Department of Physics and Astronomy, University of Oklahoma,\\
 Norman, OK 73019, USA }

\begin{abstract}
We present the effects of ellipticity of matter distribution in massive halos 
on the observation of supernovae. A pseudo elliptical Navarro-Frenk-White (NFW) 
mass model is used to calculate the introduced gain factors and observation rates 
of type Ia supernovae due to the strong lensing. We investigate how and to what
extent the ellipticity in mass distribution of the deflecting halos can affect
surveys looking for cosmologically distant supernovae. We use halo masses of 
$1.0 \times 10^{12} h^{-1} M_{\odot}$ and $1.0 \times 10^{14} h^{-1} M_{\odot}$ 
at redshifts $z_{d}=0.2$, $z_{d}=0.5$, and $z_{d}=1.0$, with ellipticities of 
up to $\epsilon=0.2$.
\end{abstract}


\section{Introduction}

Supernovae have emerged as the most promising standard candles. Due to their 
significant intrinsic brightness and relative ubiquity they can be observed in 
the local and distant universe. Observational efforts to detect high-redshift 
supernovae have proved their value as cosmological probes. The systematic study 
and observation of these faint supernovae (mainly type Ia) has been utilized 
to constrain the cosmic expansion history \citep{Goobar95, Perlmutter99, Schmidt98}.
Light emitted from any celestial object is subject to lensing by intervening objects 
while traversing the large distances involved \citep{KVB95} and the farther the
light source, the higher its chance of being significantly lensed. Apart from the
fact that gravitational lensing can limit the accuracy of luminosity distance 
measurements \citep{Perlmutter03}, it can change the observed rate of supernovae as well.
  
Studying supernovae and their rates at high redshifts provide us with much needed
information for constraining the measurements of the ellusive dark energy, as well as
understanding the cosmic star formation rate and metal enrichment at high redshifts. 
In order to observe and, hence, study the faint high-redshift supernovae, one can raise 
the chance of observation by looking through clusters of galaxies or even massive galaxies 
(see \citet{Sma02} and the references therein). These `gravitational telescopes' 
amplify the high-redshift supernovae and thereby increase the chance of their detection.
However, this boost in observation is offset by the competing effect of depletion (Fig. 1), due
to the field being spread by the deflector (amplification bias). 
For an assumed lens model 
and a given field of view it is not obvious which effect dominates the observation of
supernovae through the halo. The net result depends on the deflector and source parametrs
as well as the observational setup \citep{GunnGoo03}. 

Some research has been conducted on the feasibility of observing supernovae through
cluster of galaxies \citep[see, for instance, ][]{SaiRaySch00, GalMaoSha02, GunnGoo03}. 
These studies have not taken into account how the morphology (mainly the ellipticity) 
of these clusters as gravitational telescopes could change the expected supernova rate.
In this paper, we investigate whether introducing ellipticity into the mass distribution 
of the deflecting halos can affect the observation of supernovae. For this purpose, we use
a pseudo elliptical Navarro-Frenk-White (NFW) halo model with different values of ellipticity. 
Throughout the paper we assume the so-called concordance cosmology where $\Omega _{m}= 0.3$, 
$\Omega _{\Lambda }= 0.7$, and $h_{100}= 0.67$, with $h_{100}=H_{0}/100$ km s$^{-1}$Mpc$^{-1}$.
In $\S$ 2 we briefly go over the NFW model and show how an analytical formalism for 
a pseudo elliptical NFW mass profile can be introduced. Strong lensing by thin deflectors
as well as the way ellipticity can afffect the amplification is explained in $\S$ 3. We
present and discuss the results of our calculations in $\S$ 4.

\section{The NFW Halo Model Profile} 

\subsection{NFW Haloes}

High resolution N-body numerical simulations \citep{NavFreWhi95} 
have indicated the existence of a universal density profile for dark matter halos resulting from the 
generic dissipationless collapse of density fluctuations. This density profile does not (strongly)
depend on the mass of halo, on the power spectrum of initial fluctuations, or on the cosmological 
parameters. These halo models which are formed through
hierarchical clustering diverge with $\rho \propto r^{-1}$ near the halo center and behave as
$\rho \propto r^{-3}$ in its outer regions. Inside the virial radius, this so-called NFW halo 
profile appears to be a very good description of the mass distribution of objects spanning 9 
orders of magnitude in mass: ranging from globular clusters to massive galaxy clusters (see
\citet{WriBra00} and references therein). The NFW halo model is similar to Hernquist profile
\citep{Her90} that gives a good description of elliptical galaxy photometry. However, the two
models differ significantly at large radii, possibly due to the fact that elliptical galaxies,
countrary to the dark halos, are relatively isolated systems. 
   
The spherically symmetric NFW density profile takes the form of
\begin{equation}
\rho(r) = \frac{\delta_{c}\rho_{c}}{\displaystyle{\frac{r}{r_{s}}(1+\frac{r}{r_{s}})^{2}}}
\label{}
\end{equation}
where $\rho_{c}=[3H^{2}(z)]/(8\pi G)$ is the critical density for closure of the universe at
the redshift $z$ of the halo, $H(z)$ is the Hubble parameter at the same redshift, and $G$ is
the universal gravity constant. The scale radius $r_{s}\equiv r_{200}/c$ is the charactristic
radius of the halo where $c$ is a dimensionless number refered to as the concentration parameter,
and
\begin{equation}
\delta_{c}=\frac{200}{3}\frac{c^{3}}{\ln (1+c) - \displaystyle{\frac{c}{1+c}}}
\label{}
\end{equation}
is a charactristic overdensity for the halo. The virial radius $r_{200}$ is defined as the radius
inside which the mass density of the halo is equal to $200\rho_{c}$. It is then easy to see that
\begin{equation}
M(r_{200})\equiv M_{200}=\frac{800}{3}\rho_{c} r_{200}^{3}\, .
\label{}
\end{equation}
Therefore, NFW halos are defined by two parameters; c, and either $r_{200}$ or $M_{200}$.
\setcounter{footnote}{0}
For any spherical NFW profile with a given mass, the concentration parameter c can be 
calculated using the Fortran 77 code $\mathsf{charden.f}$ publicly available on the webpage of Julio 
Navarro\footnote{$\mathrm{http://pinot.phys.uvic.ca/ ^\sim jfn/mywebpage/jfn\_I.html}$}.

NFW halos can be shown to always produce odd number of images, as opposed to the commonly-used
singular isothermal sphere (SIS) model which produces either one or two images \citep{SchEhlFal92}. 
Although baryons are expected to isothermalize the matter distribution for halos of galaxy mass and 
below \citep{KocWhi01}, taking all of the matter in the universe in isothermal spheres is a great 
oversimplification \citep{Holz01}. It is, hence, reasonable to model halos (at least massive halos)
with NFW mass profile instead of SIS model.

\subsection{Elliptical Potential Model}

Here we present the introduced ellipticity $\epsilon$ in the circular lensing potential $\varphi(\theta)$,
assuming that angular position $\mathbf{\theta}$ can be scaled by some scale radius/angle $\theta_{s}$.
The reader is encouraged to see \citet{GolKne02} and \citet{MenBarMos03} for illuminating discussions.
We first introduce the dimensionless radial coordinates $\mathbf{x}=(x_1,x_2)=\mathbf{R}/r_s=
\mathbf{\theta}/\theta_s$ where $\mathbf{R}$ is the radial coordinate in the deflector plane, and 
$\theta_s=r_s/D_d$. Then, one can introduce the ellipticity 
in the expression of the lens potential by substituting $x_\epsilon$ for
$x$, using the following elliptical coordinate system:
\begin{equation}
\label{}
\left\lbrace
\begin{array}{lcl}
x_{1\epsilon} & = & \sqrt{a_{1\epsilon}} \, x_1 \\
x_{2\epsilon} & = & \sqrt{a_{2\epsilon}} \, x_2 \\
x_\epsilon & = & \sqrt{x_{1\epsilon}^2 +
x_{2\epsilon}^2}\ =\  \sqrt{a_{1\epsilon}x_1^2 +a_{2\epsilon}x_2^2}\\
\phi_\epsilon & = & \arctan \left(x_{2\epsilon} / x_{1\epsilon}\right)
\end{array}
\right.
\end{equation}
where $a_{1\epsilon}$ and $a_{2\epsilon}$ are the two parameters used to define the ellipticity, as
explained below. 

From the elliptical lens potential $\varphi_\epsilon(x)\equiv\varphi(x_\epsilon)$, we 
can calculatete the elliptical deflection angle (see $\S$ 3.2):
\begin{equation}
\pmb{\alpha}_\epsilon(\mathbf{x})=\left(
\begin{array}{l}
\displaystyle{\frac
{\partial\varphi_\epsilon}{\partial x_1}}=
\alpha(x_\epsilon)\sqrt{a_{1\epsilon}}\cos{\phi_\epsilon}\\
\displaystyle{\frac
{\partial\varphi_\epsilon}{\partial x_2}}=
\alpha(x_\epsilon)\sqrt{a_{2\epsilon}}\sin{\phi_\epsilon}\\
\end{array}
\right)
\label{}
\end{equation}

Notice that the expressions above hold for any definition of 
$a_{1\epsilon}$ and $a_{2\epsilon}$. Here, we follow \citet{GolKne02} who, in order to be able
to analytically derive the convergence and shear, chose the following elliptical parameters:
\begin{eqnarray}
\label{}
a_{1\epsilon} & = & 1-\epsilon  \\ 
a_{2\epsilon} & = & 1+\epsilon
\end{eqnarray}
which for small values of ellipticity $\epsilon$ results in the same ellipticity along the $x_1$
as the standard elliptical model of
\begin{eqnarray}
\label{}
a_{1\epsilon} & = & 1-\epsilon     \\
a_{2\epsilon} & = & 1/(1-\epsilon)
\end{eqnarray}
with $\epsilon=1-b/a$, where $a$ and $b$ are the semi-major and semi-minor axis of the projected
elliptic potential, respectively.

\section{Gravitational Lensing: a Reminder}

\subsection{General Formalism}
In the thin-lens approximation, we define $\mathfrak{z}$ as the optical axis and $\Phi(R,\mathfrak{z})$
as the 3-dimensional Newtonian potential, with $r=\sqrt{R^2+\mathfrak{z}^2}$. The so-called reduced
2-dimensional potential which is defined in the deflector plane is given by
\begin{equation}
\varphi(\mathbf{\theta})=\frac{2}{\mathfrak{c}^2}\frac{D_{ds}}{D_{d}D_{s}}
\int\limits_{-\infty}^{+\infty} \Phi(D_{d}\,\theta,\mathfrak{z})\,d\mathfrak{z}
\label{}
\end{equation}
\citep{SchEhlFal92} where $\mathfrak{c}$ is the speed of light, and $\mathbf{\theta}=(\theta_1,\theta_2)$ 
is the angular position in the image plane. $D_{d}$, $D_{s}$, and $D_{ds}$ are
angular distances of observer-deflector, observer-source, and deflector-source, respectively.
The deflection angle $\mathbf{\alpha}$, convergence $\kappa$ and the shear $\gamma$ are given by
the following set of equations:
\begin{equation}
\left\lbrace
\begin{array}{l}
\pmb{\alpha}(\theta)=\mathbf{\nabla}_{\mathbf{\theta}}\varphi(\theta)\\
\kappa(\theta)=\displaystyle{\frac{1}{2}\left(\frac{\partial^2\varphi}{\partial\theta_1^2}
+\frac{\partial^2\varphi}{\partial\theta_2^2}\right)}\\
\gamma^2(\theta)=\|\mathbf{\gamma}(\theta)\|^2=\displaystyle{
\frac{1}{4}\left(\frac{\partial^2\varphi}{\partial\theta_1^2}
-\frac{\partial^2\varphi}{\partial\theta_2^2}\right)^2+
\left(\frac{\partial^2\varphi}{\partial\theta_1\partial\theta_2}\right)^2}.
\end{array}
\right.
\label{}
\end{equation} 
The lensing equation then reads:
\begin{equation}
\mathbf{\beta}=\mathbf{\theta}-\mathbf{\alpha}=
\mathbf{\theta}-{\mathbf{\nabla}}_{\theta}\varphi(\theta) 
\label{}
\end{equation}
where $\mathbf{\beta}=(\beta_1, \beta_2)$ is the angular location of the source. The amplification $amp$ of
a point image formed at $\theta$ is:
\begin{equation}
amp(\theta) = \frac{1}{(1-\kappa)^2-\gamma^2}
\label{}
\end{equation}

To calculate the angular distances in our work, we use the solution to the Lam\'{e}
equation for the distance-redshift equation in a partially filled beam
Friedmann-Lema\^{i}tre-Robertson-Walker (FLRW) cosmology. For a filled-beam flat
FLRW cosmology, the angular distance $D$ as a function of redshift $z$ is  
\begin{equation}
D(z)=\frac{2\mathfrak{c}z}{(1+z)H_{0}\left(g(z)\right)^{1/2}}\, _{2}F_{1}\left( \frac{1}{6},\frac{1}{2};
\frac{7}{6}, -\left[ \frac{(\Omega_{m}^{2}\Omega_{\Lambda})^{1/3}z^{2}}{g(z)}\right]^{3}\right) 
\end{equation} 
where
\begin{equation}
g(z) \equiv 2\sqrt{1+\Omega_{m}z(3+3z+z^{2})}+2+\Omega_{m}z(3+z) \, .
\label{}
\end{equation}
See \citet{Kant03} for more detail.

\subsection{Lensing Parameters of Spherically symmetric NFW Model}

Several authors have developed the lensing equations for the ordinary, spherical NFW halos
\citep[e.g.][]{Bart96, WriBra00, GolKne02}. Following $\S$ 2.2 we can introduce a dimensionless radial coordinate
in the lens plane $\mathbf{x}=(x_1,x_2)=\mathbf{R}/r_s=\mathbf{\theta}/\theta_s$ 
where $\theta_s=r_s/D_d$. The surface mass density then becomes
\begin{equation}
\Sigma(x)=\int\limits_{-\infty}^{+\infty}\rho(r_s\,x,z)dz=2\delta_c \rho_c r_s F(x)
\label{}
\end{equation}
with
\begin{equation}
F(x)=
\begin{cases}
\displaystyle{\frac{1}{x^2-1}\left(1-\frac{1}{\sqrt{1-x^2}} \mathrm{arcch}\frac{1}{x}\right)} & (x<1) \\
\displaystyle{\frac{1}{3} }                                                                   & (x=1) \\
\displaystyle{\frac{1}{x^2-1}\left(1-\frac{1}{\sqrt{x^2-1}} \arccos{\frac{1}{x}}\right)}      & (x>1)
\end{cases}
\end{equation}
and the mean surface density inside the radius $x$ can be written as
\begin{equation}
\overline{\Sigma}(x)=\displaystyle{\frac{1}{\pi x^2}\int\limits_{0}^{x}
2\pi x\Sigma(x)dx=4\delta_c \rho_c r_s \frac{g(x)}{x^2}}
\label{}
\end{equation}
with
\begin{equation}
g(x)=
\begin{cases}
\displaystyle{\ln{\frac{x}{2}}+\frac{1}{\sqrt{1-x^2}} \mathrm{arcch} \frac{1}{x}}  & (x<1) \\
\displaystyle{1+\ln{\frac{1}{2}}}                                                  & (x=1) \\
\displaystyle{\ln{\frac{x}{2}}+\frac{1}{\sqrt{x^2-1}}\arccos{\frac{1}{x}}}         & (x>1)
\end{cases}
\end{equation}
(see \citet{GolKne02}).

The deflection angle $\mathbf{\alpha}$, convergence $\kappa$ and shear $\gamma$ turn 
out as
\begin{equation}
\left\lbrace
\begin{array}{llcl}
\mathbf{\alpha}(x)& = & \theta \, 
\displaystyle{\frac{\overline{\Sigma}(x)}{\Sigma_\mathrm{crit}}}
& = 4\kappa_s \, 
\displaystyle{\frac{\theta}{x^2}}g(x)\mathbf{e}_x \\
\kappa(x) & = & 
\displaystyle{\frac{\Sigma(x)}{\Sigma_\mathrm{crit}}}
& = 2\kappa_s \, F(x)\\
\gamma(x) & = &
\displaystyle{\frac{\overline{\Sigma}(x)-\Sigma(x)}{\Sigma_\mathrm{crit}}}
& = 2\kappa_s \left({\displaystyle{\frac{2g(x)}{x^2}-F(x)}}\right)
\end{array}
\right.
\label{}
\end{equation}
where $\kappa_s= \delta_c \rho_c r_s\Sigma_\mathrm{crit}^{-1}$, with $\Sigma_\mathrm{crit}
\equiv \mathfrak{c}^{2} D_{s} / (4 \pi G D_{d} D_{ds})$.

By integrating the deflection angle, the potential $\varphi(x)$ can be found:
\begin{equation}
\varphi(x)=2\kappa_s\theta_s^2\,h(x)
\label{}
\end{equation}
with
\begin{equation}
h(x)=
\begin{cases}
\displaystyle{\ln^2{\frac{x}{2}}-\mathrm{arcch}^2
\frac{1}{x}} & (x<1)\\
\displaystyle{\ln^2{\frac{x}{2}}+\arccos^2{\frac{1}{x}}}
 & (x\ge1)
\end{cases}
\end{equation}

\subsection{Lensing Parameters of Pseudo Elliptical NFW Model}
	
For the particular choice of $\epsilon$ in $\S$ 2.2, the corresponding convergence and shear can
be calculated:
\begin{eqnarray}
\kappa_\epsilon(\boldmath{x}) & = & \displaystyle{\frac{1}{2\theta_s^2}\left(
\frac{\partial^2\varphi_\epsilon}{\partial x_1^2}
+\frac{\partial^2\varphi_\epsilon}{\partial x_2^2}\right)}             \nonumber \\
& = & \kappa(\boldmath{x}_\epsilon)+\frac{\epsilon}{2\theta_s^2}\left(
\frac{\partial^2\varphi(x_\epsilon)}{\partial x_{2\epsilon}^2}
-\frac{\partial^2\varphi(x_\epsilon)}{\partial x_{1\epsilon}^2}\right) \nonumber \\
& = & \kappa(\boldmath{x}_\epsilon)+\epsilon\cos{2\phi_\epsilon}\,\gamma(\boldmath{x}_\epsilon).
\label{}
\end{eqnarray}
and
\begin{eqnarray}
\gamma_\epsilon^2(\boldmath{x}) & = & \frac{1}{4\theta_s^4}\left\lbrace\left(\frac{\partial^2\varphi_\epsilon}
{\partial x_1^2}-\frac{\partial^2\varphi_\epsilon}{\partial x_2^2}\right)^2+
\left(2\frac{\partial^2\varphi_\epsilon}{\partial x_1\partial x_2}\right)^2\right\rbrace     \nonumber \\
& = & \gamma^2(\boldmath{x}_\epsilon) + 2\epsilon\cos{2\phi_\epsilon}\gamma(\vec{x}_\epsilon)
\kappa(\boldmath{x}_\epsilon) + \epsilon^2(\kappa^2(\boldmath{x}_\epsilon)-\cos^2{2\phi_\epsilon}
\gamma^2(\boldmath{x}_\epsilon)).
\end{eqnarray}

Also, the elliptic projected mass density reads:
\begin{equation}
\Sigma_\epsilon(\mathbf{x})=\Sigma(\mathbf{x}_\epsilon)+\epsilon\cos{2\phi_\epsilon}
(\overline{\Sigma}(\mathbf{x}_\epsilon)-\Sigma(\mathbf{x}_\epsilon)).
\label{}
\end{equation} 

The lensing equation now becomes (see the appendix):
\begin{equation}
\left\lbrace
\begin{array}{lcl}
\beta_{1} & = & \displaystyle{\theta_{s} x_{1} \left( 1-4k_{s}\epsilon_{1} \frac{g(x_{\epsilon})}{x_{\epsilon}^{2}} \right)}  \\
\beta_{2} & = & \displaystyle{\theta_{s} x_{2} \left( 1-4k_{s}\epsilon_{2} \frac{g(x_{\epsilon})}{x_{\epsilon}^{2}} \right)}
\end{array}
\right.
\end{equation}
and as one expects, the amplification $amp$ reads:
\begin{equation}
amp(\mathbf{x})=\frac{1}{(1-\kappa_\epsilon(\mathbf{x}))^{2} - \gamma_\epsilon^2(\mathbf{x})}
\end{equation}

It can be shown that ellipticities beyond $\epsilon=0.2$ result in unrealistic `peanut' shaped projected
densities, hence in this work we focus on lower values of $\epsilon$.
Figure 2 shows the multiple images produced by a $1.0 \times 10^{14}  h^{-1} M_{\odot}$ halo with ellipticity
$\epsilon = 0.1$ (courtesy of Golse \& Kneib). 
Dashed lines are the contours with constant surface density $\Sigma_{\epsilon}$ and the solid
lines are the critical and caustic lines. Redshifts of source and deflector are 0.2 and 1.0, respectively.

\section{The Method}

The main reason for studying supernovae magnified by gravitational lensing is to investigate the chance of
observing supernovae too faint to be observed in the absence of lensing, which is usually the case for
cosmologically distant supernovae, specifically type Ia's.
To calculate the observed rate of type Ia supernovae we use the result of predicted rates by \citet{DahFra99} 
for a hierarchical star formation rate model with a charactristic time of $\tau=1$ Gyr (Fig. 3), which limits
our calculation to the redshift depth of $z_{Max} = 5$.

In order for a supernova to be detected, its apparent magnitude $m$ should not exceed the limiting magnitude of
the survey $m_{limit}$. Using the definitions of the apparent magnitude and amplification, we get:
\begin{equation}
m_{amp}=m_{o}+2.5\log (\left|\left(1-\kappa \right)^{2}-\gamma ^{2}\right|)\
\end{equation}
in which, $m_{amp}$ is the observed magnitude, and $m_{o}$ is the apparent magnitude of the supernova in the 
absence of the lensing. We can further write $m_{o}$ in terms of the absolute magnitude $M_{abs}$ of the
supernova and rewrite the detection criterion as
\begin{equation}
\left((1-\kappa )^{2}-\gamma ^{2}\right)D_{L}^{2}(z_{s})\leqslant 10^{\left(\frac{m_{limit}-M_{abs}+5}{2.5}\right)}
\end{equation} 
where $D_{L}(z_{s})$ is the luminosity distance of the supernova at redshift $z_{s}$. The absolute magnitude of 
type Ia SNe has a very narrow Gaussian distribution around $M_{abs}=-19.16$ at
a confidence level of 89\% \citep{Rich02} . Here, we assume that the supernova is detected as soon as its absolute 
magnitude becomes brighter than $M_{abs}=-18$. 

We take the deflecting halo to be at redshifts $z_{s}$ = 0.2, 0.5, and 1.0, and with virial 
masses of $m_{d1}=1.0 \times 10^{12} h^{-1} M_{\odot}$ and $m_{d2}=1.0 \times 10^{14}M_{\odot} h^{-1}$. 
Concentration parameter c, overdensity $\delta_{c}$, and virial radius $r_{200}$ (in units of $Kpc h^{-1}$) 
for each case are given in Table 1. 

\begin{table}[b]
 \begin{center}
  \begin{tabular}{| c || c | c | c || c | c | c | c |}
    \hline
         $\mathbf{m_{d}}$ & \multicolumn{3}{c ||}{$1.0 \times 10^{12} h^{-1} M_{\odot}$} & 
         \multicolumn{3}{c |}{$1.0 \times 10^{14} h^{-1} M_{\odot}$}  \\  
    \hline \hline    
         $\mathbf{z_{d}}$ & $\mathbf{r_{200}}$ & $\pmb{\delta_{c}}$ & $\mathbf{c}$ & 
         $\mathbf{r_{200}}$ & $\pmb{\delta_{c}}$ & $\mathbf{c}$  \\   
    \hline 
         0.2 & 152.61 & 38468.6 & 9.40 & 708.36 & 15741.7 & 6.46  \\            
    \hline 
         0.5 & 136.24 & 33096.0 & 8.83 & 632.38 & 14426.7 & 6.22  \\         
    \hline      
         1.0 & 111.79 & 25118.2 & 7.87 & 518.88 & 12086.5 & 5.77  \\           
    \hline       
   \end{tabular}  
   \caption{NFW halo parameters for the two halo masses $m_{d}$ at the given redshifts $z_{d}$ used in the paper.}      
 \end{center}
 \label{T1}
\end{table}  
      
The field of view is taken to be the spatial angle subtending the virial area of the halo. By breaking the projected
halo into pixels with the angular size of $\delta x_{1}$ and $\delta x_{2}$ (which are taken to be smaller than the 
angular resolution of the observation, Figure 4), we calculate the amplification across the halo and hence, find the 
number of observable supernovae in redshift shells with the width of $\delta z = 0.05$. We find the corresponding
(spatial angular) element $\delta\beta_{1} \times \delta\beta_{2}$ in the area behind the halo 
(in redshift space) where the supernovae are bright enough to be detected. Assuming we can 
arbitrarily minimize $\delta x_{1}$ and $\delta x_{2}$ , we have 
\begin{equation}
\delta\beta_{1} \times \delta\beta_{2} = 
\left| \frac{\partial \pmb{\beta}}{\partial \pmb{x}} \right|
\delta x_{1} \times \delta x_{2}
\end{equation}
where $\left| \frac{\partial \pmb{\beta}}{\partial \pmb{x}} \right|$ is the determinant of Jaccobian matrix.

The reader can refer to the appendix for the derivation of the Jaccobian. The gain factor, defined as the ratio
of the number of observable lensed supernovae over the number of observable supernovae in the absence of lensing
($N_{lensed}/N_{No Lensing}$) can be calculated by integrating over the predicted rates of both cases across
the whole observable area (Fig. 1) for any given lensing configuration, considering the ellipticity $\epsilon$.

\section{Results and Discussion}
 
First, we consider the effect of ellipticity in the number rate of SN Ia in every redshift bin $\delta z=0.05$.
Upper panels of Figures 5 ($m_{d1}$) and 6 ($m_{d2}$)
show the number of expected supernovae per year occuring in the redshift bins. We present the results for  
$\epsilon=0.0$ and $\epsilon=0.2$ with the deflecting halo at redshifts $z_{d}=$ 0.2, 0.5, and 1.0. The survey magnitude
is assumed to be $m_{lim} = 27$. The number rate peaks at around $z=1.3$ as expected (see Fig. 3) and dies off rapidly
beyond that. It can be seen that the farther the deflector, the slightly higher the slope of the curves up to $z=1.3$ as a 
result of higher number of supernovae observed in front of the deflector. 

Middle and lower panels in Figures 5 and 6 show the cumulative number rates and the gains, respectively. The dominance of 
amplification bias as a result of the narrowing of the field in a region immediately behind the deflectors at the assumed redshifts 
is clear, as the gains fall below 1. Beyond that region amplification takes over and more (lensed) supernovae are observed.

In the absence of an intercepting halo, the number rate of the survey drops to zero at the redshift limit of the survey.
With the deflecting halo present, the observed rate goes to zero at a higher redshift. This can be seen in
figures 7 ($m_{d1}$) and 8 ($m_{d2}$) where the deflector is at redshift $z_{d}=0.5$ and the survey magnitude limit is 
$m_{lim} = 27$. The three upper panels depict the expected rates for lensing and no-lensing sccenarios for ellipticities
$\epsilon=$ 0.0, 0.1, and 0.2. The number rates per redshift bin (left) and the cumulative rate (right) are given. The 
reader can readily notice the effect of bias behind the halo. With the galactic size halo $m_{d1}$, the survey can detect 
supernovae up to redshift $z \sim 3$ (Fig. 7). This limit increases to $z \sim 5$ (Fig. 8) for the cluster-size halo $m_{d2}$.

The lowest panel in these two figures show the relative difference of the cases with $\epsilon = 0.1$ and $\epsilon = 0.2$
with respect to $\epsilon = 0.0$. The 2 curves do not show significant difference for the redshift bins in front of the
halo. In the regime behind the halo, the difference becomes remarkable: it increases up to redshift $z=1.4$ for $m_{d1}$ and
$z=1.7$ for $m_{d2}$. The difference doesn't vary remarkably beyond the maximum point. 
 
To further see how ellipticity changes the expected rate of observed supernovae we put the result of our calculations
for different ellipticities for a given range of magnitude limits on the same plot. Figure 9 shows the number rate of 
observed type Ia supernovae (upper panel) for ellipticities $\epsilon = 0.1$ and $\epsilon = 0.2$ together with their
relative difference with respect to the case with no ellipticity (lower panel). Both halo masses, $m_{d1}$ and $m_{d2}$ 
are at redshift $z_{d}=0.2$. Figures 10 and 11 show the results of the same calculations with halos at redshifts
$z=0.5$ and $z=1.0$, respectively. The number rates in each figure increase smoothly up to the magnitude limit at which
the survey is deep enough to detect the supernovae as far as the halo itself, e.g, $m_{lim}=22.4$ for a concordance cosmology of 
($\Omega_{m}, \omega_{\Lambda}, h_{100}$) = (0.3, 0.7, 0.67). From that point on the rates increase very rapidly as the magnitude
limit goes up. That is caused by the halo lensing and hence amplifying the supernovae which would otherwise be too dim to
be observed. The relative differences depicted in these figures show that even at a magnitude limit of 25, effect of ellipticity 
cannot be ignored as it significantly changes the number/percentage of the observed supernovae; for instance, the relative 
difference for $\epsilon=0.2$ with deflecting halo $m_{d2}$ at redshift $z=1.0$ (Fig. 11) exceeds 9\% for the magnitude limit
of $m_{lim}=27$.

\section{Conclusion}

Aiming behind massive halos seem to be a good way to enhance the high-redshift supernovae surveys. The cumulative gains 
of such surveys seem insignificant at low redshifts ($z_{s}<0.2$) but the results are remarkable at higher redshifts.
For deep observations where $m_{lim}>25$, the geometry of the intervening halo cannot be ignored.
We have shown that introducing ellipticity in the (gravitational potential of) the mass distribution
of a deflecting halo (here, for a galactic halo of mass $1.0 \times 10^{12} M_{\odot} h^{-1}$ as well as a middle-size 
cluster of galaxies with a mass of $1.0 \times 10^{14} M_{\odot} h^{-1}$) can affect the rate of observed supernovae
by a few percent. It was shown that the farther the supernova survey probes, the more significant the effects of
introduced ellipticity become.

It should be noted that this work does not involve a broad range of mass profiles for the halos (although we specify that
the survey is limited to the virial area of the halo), nor does it address the much needed k-correction. Our calculations are 
actually an oversimplification due to the fact that a large, massive halo like a galaxy cluster has substructure 
which consists of the member galaxies, as well as large clouds of gas. A more sophisticated lens model with ellipticity 
should be employed to calculate the number rate of observed supernovae. 
 
\acknowledgements{}
 
The author wishes to thank D. Branch for enlightening discussions and suggestions. The author would 
also like to thank C. Gunnarsson and A. Goobar for generously offering him their data set on the rate of SNe Ia, 
and J. Navarro for allowing him to use the NFW code.
This work was in part supported by NSF grant AST0204771 and NASA grant NNG04GD36G.

\appendix
\section{Appendix}

Here we derive the lensing equation for an elliptical NFW halo with
ellipticity of $\epsilon$ introduced in its 2-dimensional potential, and proceed 
to calculate Jaccobian $\frac{\partial \pmb{\beta}}{\partial \pmb{x}}$ needed to get the spatial angular element
$\delta\beta_{1} \times \delta\beta_{2}$ in the source frame.  

\subsubsection*{Lensing Equation}

Introducing the dimensionless coordinate system
$\mathbf{x}=(x_1,x_2)=\mathbf{R}/r_s=\mathbf{\theta}/\theta_s$, the
lensing equation becomes
\begin{equation}
\left\lbrace
\begin{array}{lcl}
\beta_{1} & = & \theta_{s} x_{1} - \alpha_{1}\left[ x_{1},x_{2} \right]  \\
\beta_{2} & = & \theta_{s} x_{2} - \alpha_{2}\left[ x_{1},x_{2} \right] 
\end{array}
\right.
\end{equation}
Given the elliptical deflection angle of
\begin{equation}
\mathbf{\alpha}_\epsilon(\mathbf{x})=\left(
\begin{array}{l}
\displaystyle{\frac
{\partial\varphi_\epsilon}{\partial x_1}}=
\alpha(x_\epsilon)\sqrt{a_{1\epsilon}}\cos{\phi_\epsilon}\\
\displaystyle{\frac
{\partial\varphi_\epsilon}{\partial x_2}}=
\alpha(x_\epsilon)\sqrt{a_{2\epsilon}}\sin{\phi_\epsilon}\\
\end{array}
\right)
\label{}
\end{equation}
and the deflection angle of $\alpha$ as
\begin{equation}
\pmb{\alpha}(x)  \,= \,  \theta \, 
\displaystyle{\frac{\overline{\Sigma}(x)}{\Sigma_\mathrm{crit}}}
\, = \, 4\kappa_s \, \displaystyle{\frac{\theta}{x^2}}g(x)\mathbf{e}_x 
\end{equation}
the lensing equation now reads
\begin{equation}
\left\lbrace
\begin{array}{lcl}
\beta_{1} & = & \theta_{s} x_{1} \left( 1-4k_{s}\epsilon_{1} \displaystyle{\frac{g(x_{\epsilon})}{x_{\epsilon}^{2}}} \right)  \\
\beta_{2} & = & \theta_{s} x_{2} \left( 1-4k_{s}\epsilon_{2} \displaystyle{\frac{g(x_{\epsilon})}{x_{\epsilon}^{2}}} \right)
\end{array}
\right.
\label{b:x}
\end{equation}

\subsubsection*{Jaccobian}

To calculate spatial angular element $\delta\beta_{1} \times \delta\beta_{2}$ 
we use the Jaccobian equation
\begin{equation}
\delta\beta_{1} \times \delta\beta_{2} 
\, =\,  
\left| \frac{\partial \pmb{\beta}}{\partial \pmb{x}} \right|
\delta x_{1} \times \delta x_{2} 
\, =\, 
\left| \frac{\partial \beta_{1}}{\partial x_{1}}.\frac{\partial \beta_{2}}{\partial x_{2}} - 
\frac{\partial \beta_{1}}{\partial x_{2}}.\frac{\partial \beta_{2}}{\partial x_{1}}  \right|
\delta x_{1} \times \delta x_{2}   
\end{equation}
Given Eq. \ref{b:x}, we get:
\begin{equation}
\label{}
\left\lbrace
\begin{array}{lcl}
\displaystyle{\frac{\partial \beta_{1}}{\partial x_{1}}} &=& \theta_{s}\left(1-4k_{s}a_{1\epsilon}G(x_{\epsilon})\right)
      \theta_{s} x_{1} \left(1-4k_{s}a_{1\epsilon}\displaystyle{\frac{\partial G(x_{\epsilon})}{\partial x_{1}}}   \right)  \\
\displaystyle{\frac{\partial \beta_{2}}{\partial x_{1}}} &=& 
      \theta_{s} x_{2} \left(1-4k_{s}a_{2\epsilon}\displaystyle{\frac{\partial G(x_{\epsilon})}{\partial x_{1}}}   \right)  \\
\displaystyle{\frac{\partial \beta_{1}}{\partial x_{2}}} &=& 
      \theta_{s} x_{1} \left(1-4k_{s}a_{1\epsilon}\displaystyle{\frac{\partial G(x_{\epsilon})}{\partial x_{2}}}   \right)  \\
\displaystyle{\frac{\partial \beta_{2}}{\partial x_{2}}} &=& \theta_{s}\left(1-4k_{s}a_{2\epsilon}G(x_{\epsilon})\right)
      \theta_{s} x_{2} \left(1-4k_{s}a_{2\epsilon}\displaystyle{\frac{\partial G(x_{\epsilon})}{\partial x_{2}}}   \right)
\end{array}
\right.
\end{equation}
where function $G$ is defined as
\begin{equation}
G(x_{\epsilon}) \equiv \frac{g(x_{\epsilon})}{x^{2}_{\epsilon}}
\end{equation}
and
\begin{equation}
\frac{g(x_{\epsilon})}{x^{2}_{\epsilon}} = 
\begin{cases}
\displaystyle{\frac{x_{\epsilon}}{\left(1-x^{2}_{\epsilon}\right)^{\frac{3}{2}}} 
   \mathrm{arcch}\frac{1}{x_{\epsilon}} - \frac{\left(1+x^{2}_{\epsilon}\right)}{2x_{\epsilon}\left(1-x^{2}_{\epsilon}\right)} }
   & (x_{\epsilon}<1) \\
\displaystyle{-\frac{1}{6} }
   & (x_{\epsilon}=1) \\
\displaystyle{\frac{\left(1+x^{2}_{\epsilon}\right)}{2x_{\epsilon}\left(1-x^{2}_{\epsilon}\right)} -
   \frac{x_{\epsilon}}{\left(1-x^{2}_{\epsilon}\right)^{\frac{3}{2}}} \mathrm{arccos}\frac{1}{x_{\epsilon} } }
   & (x_{\epsilon}>1) .
\end{cases}
\end{equation}

\clearpage

\figcaption{}
Schematic picture of the lensing configuration by a deflecting halo. $z_{halo}$ 
is the redshift of the halo and $z_{limit}$ is the redshift corresponding to the limiting
magnitude $m_{limit}$. The shaded area shows the volume where SNe are bright enough
to be observed.
 
\label{f1}

\figcaption{}
Multiple images produced by a $1.0 \times 10^{14}  h^{-1} M_{\odot}$ halo with ellipticity
$\epsilon = 0.1$. Dashed lines are the contours with constant surface density and the solid
lines are the critical and caustic lines. Redshifts of source and deflector are 0.2 and 1.0,
respectively (courtesy of Golse \& Kneib).
\label{f2}

\figcaption{}
Rates of type Ia supernovae in intervals of $\delta z = 0.05$. Dilution factor of $1+z$ is
taken into account (courtesy of Gunnarsson \& Goobar).
\label{f3}

\figcaption{}
This figure shows how the projected deflector is `pixellated' in order to calculate the 
observable area behind the halo. Each pixel has dimensions of $\delta \omega \times \delta \omega$
with $\delta \omega$ being smaller than the angular resolution of the observation.
\label{f4}

\figcaption{}
Rates of observed supernovae Ia per redshift bin $\delta_{z}=0.05$ (upper panel), cumulative
rate (middle panel), and the lensing gain (lower panel) for a deflecting halo of mass
$m_{d}=1.0 \times 10^{12}  h^{-1} M_{\odot}$ at redshifts of $z_{d}=0.2$, 0.5, and 1.0 with
ellipticities $\epsilon=$ 0.1 and 0.2.
\label{f5}

\figcaption{}
Same as Figure 5, with $m_{d}=1.0 \times 10^{14} M_{\odot} h^{-1}$.
\label{f6}

\figcaption{}
In this picture the 3 upper panels show observed rates of lensed (solid line) and unlensed (dash line)
for three different ellipticies $\epsilon=$ 0.0, 0.1, and 0.2. The deflecting halo has a mass of
$m_{d}=1.0 \times 10^{12}  h^{-1} M_{\odot}$ and is located at redshift $z_{s}=$ 0.5. The lowermost panel
depicts the relative difference of $\epsilon=$ 0.1 (solid line) and $\epsilon=$ 0.2 (dash line) with
respect to $\epsilon=$ 0.0.
\label{f7}

\figcaption{}
Same as Figure 7, with $m_{d}=1.0 \times 10^{14} M_{\odot} h^{-1}$.
\label{f8}

\figcaption{}
Rates of observed supernovae Ia as a function of survey magnitude limit $m$ (upper panel). Result sare 
shown for halo masses $m_{d}=1.0 \times 10^{12}  h^{-1} M_{\odot}$ and $m_{d}=1.0 \times 10^{14}  h^{-1} M_{\odot}$ 
with ellipticities $\epsilon =$ 0.1 and 0.2. The halo is at redshift $z_{d}=0.2$. The relative difference
of cases with $\epsilon =$ 0.1 and 0.2 with respect to $\epsilon =$ 0.0 is given in the lower panel.
\label{f9}

\figcaption{}
Same as Fig. 9 with $z_{d}=0.5$.
\label{f10}

\figcaption{}
Same as Fig. 9 with $z_{d}=1.0$.
\label{f11}

\clearpage

\plotone{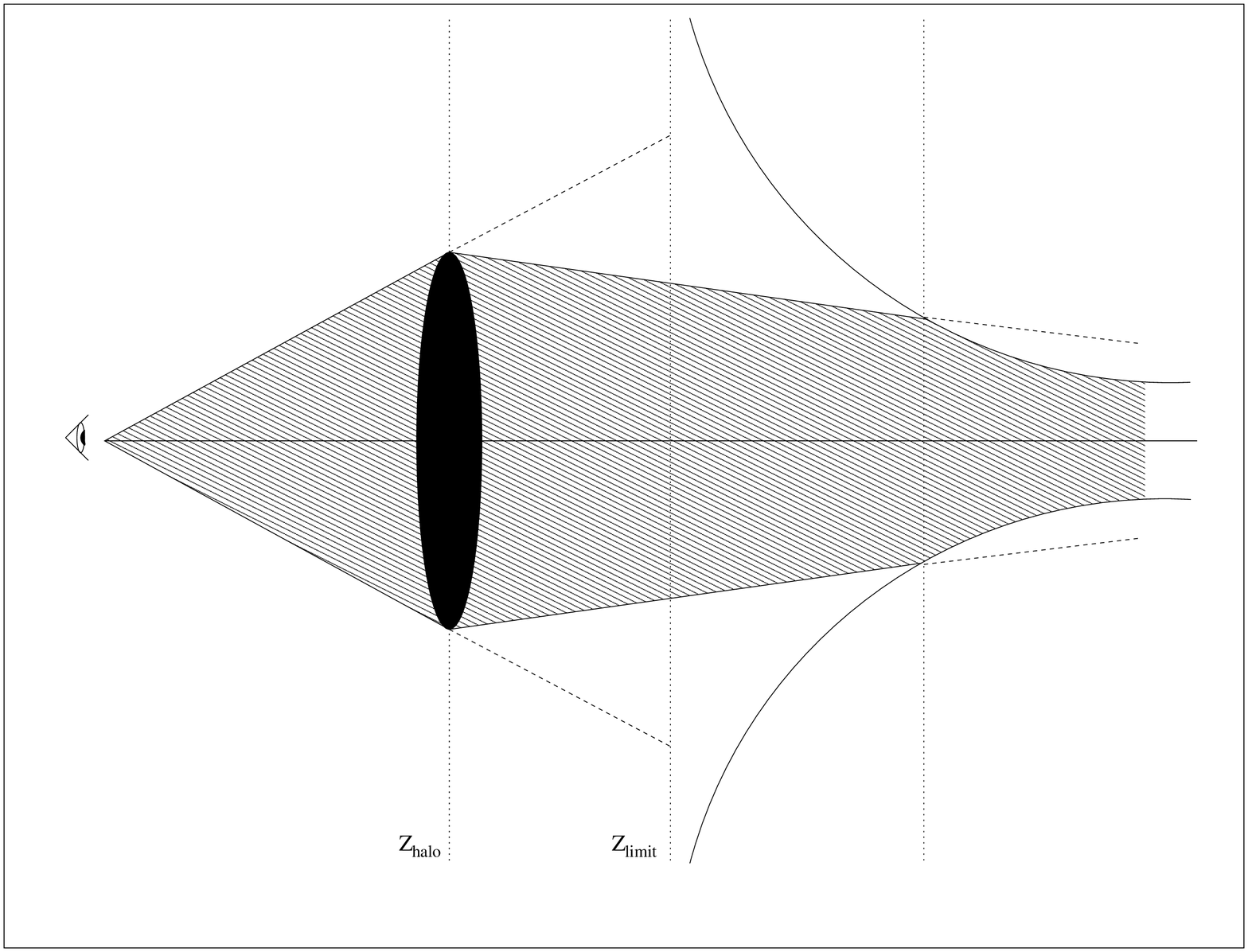}
\begin{center} Figure 1 \end{center}
\eject

\centering
\plotone{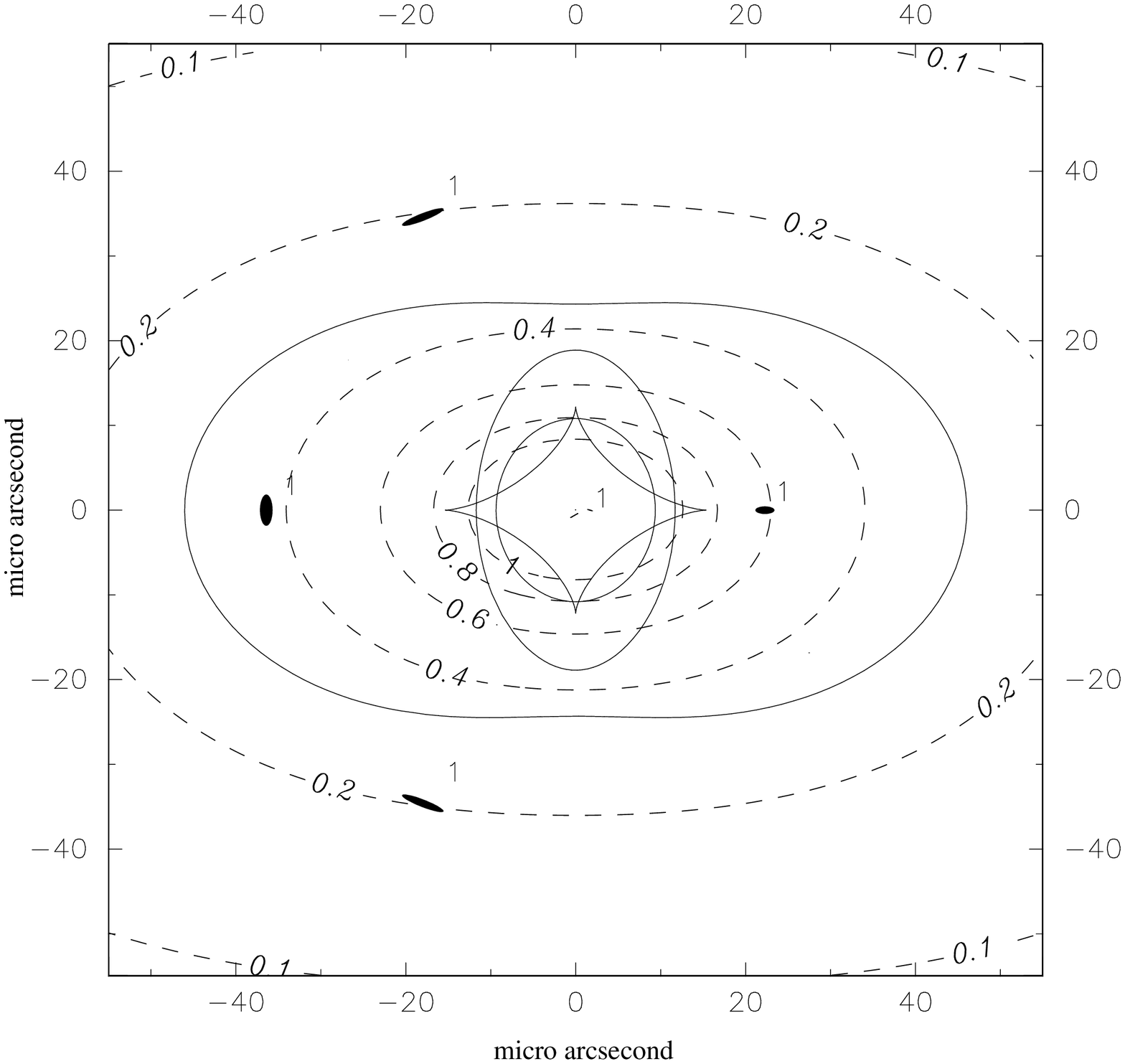}
\begin{center} Figure 2 \end{center} 
\eject

\plotone{f3.eps}
\begin{center} Figure 3 \end{center}
\eject

\plotone{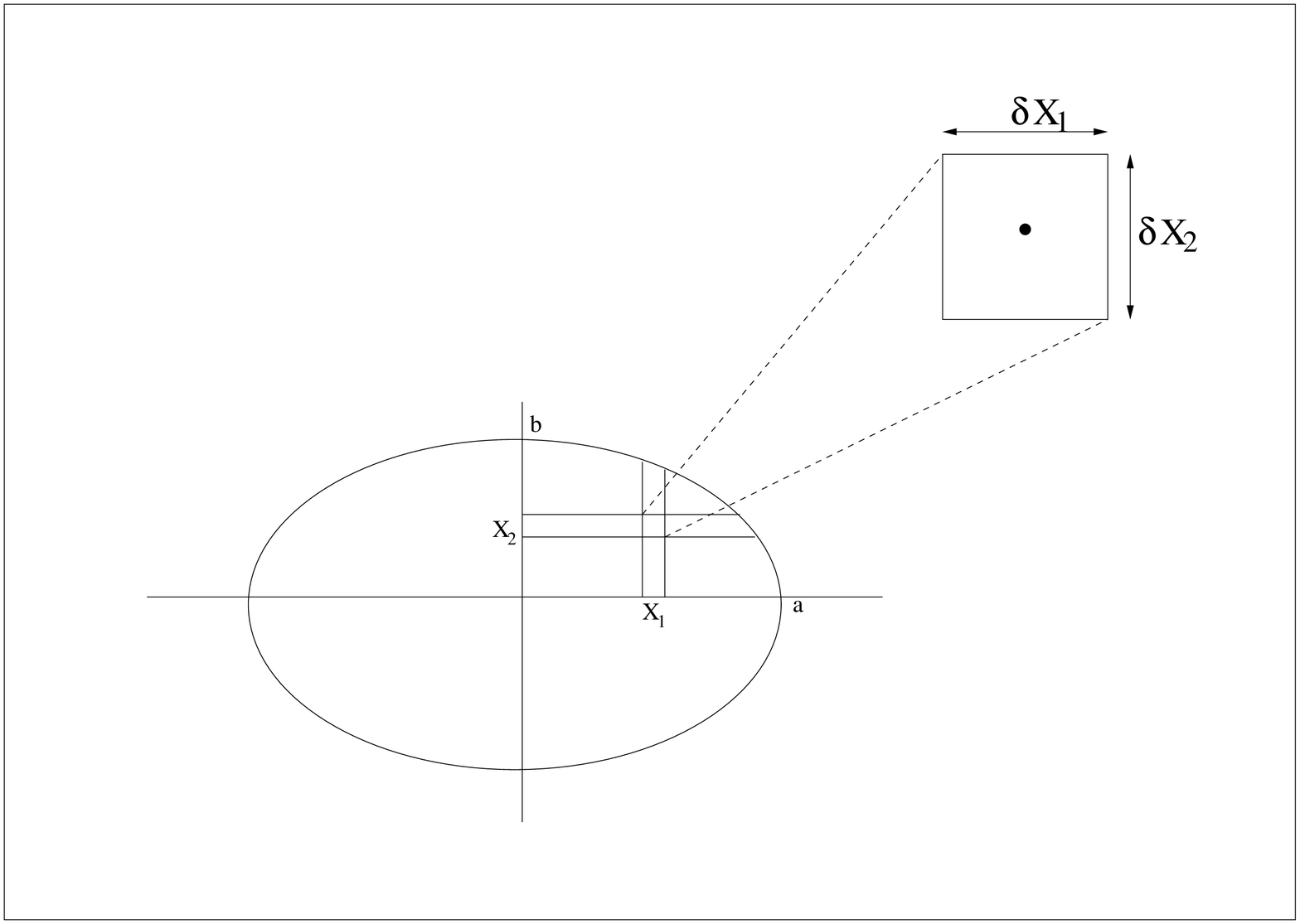}
\begin{center} Figure 4 \end{center}
\eject

\plotone{f5.eps}
\begin{center} Figure 5 \end{center}
\eject

\plotone{f6.eps}
\begin{center} Figure 6 \end{center}
\eject

\epsscale{0.85}
\plotone{f7.eps}
\begin{center} Figure 7 \end{center}
\eject

\plotone{f8.eps}
\begin{center} Figure 8 \end{center}
\eject

\epsscale{1}
\plotone{f9.eps}
\begin{center} Figure 9 \end{center}
\eject

\plotone{f10.eps}
\begin{center} Figure 10 \end{center}
\eject

\plotone{f11.eps}
\begin{center} Figure 11 \end{center} 
\eject

\end {document}